\documentclass{aa}  

\usepackage{graphicx}
\usepackage{txfonts}
\usepackage{lipsum}
\usepackage{subcaption}         
\usepackage{lscape}             
\usepackage{placeins}           

\usepackage[dvipsnames]{xcolor} 
\usepackage{multirow} 
\usepackage{multicol}
\usepackage{soul} 

\usepackage{enumitem}
\usepackage{hyperref}
\hypersetup{
    colorlinks=true,
    linkcolor=blue,
    citecolor=blue,
    filecolor=magenta,      
    urlcolor=blue,
}

\newcommand{\dd}{\text{d}}

\newcommand{\low}[1]{_\text{#1}}

\newcommand{\ME}{M_\oplus}

\begin{document}

\title{The evolution and internal structure of Neptunes and sub-Neptunes II}

\subtitle{Convective mixing and thermal conductivity}

\author{Mark Eberlein\inst{1}\thanks{Corresponding author: mark.eberlein@uzh.ch}
    \and Ravit Helled\inst{1}
    }

\institute{Department of Astrophysics, University of Zurich, Winterthurerstrasse 190, 8057 Zurich, Switzerland}

\date{Submitted: 20 February 2026, accepted 27 May 2026}

\abstract
{Sub-Neptunes and Neptunes are often modeled with distinct, fully convective layers. Yet, there are several arguments for compositions gradients that can inhibit convection. In these regions, energy transport depends on the thermal conductivity and radiative opacity.}
{We compare three thermal conductivity models and investigate  their impact on planetary evolution accounting for the possibility of convective mixing eroding composition gradients.} 
{Using a modified version of MESA, we model the evolution of planets with masses of $M\low{p}=5,\,10,\,15\,\ME$ and three initial entropies. We implement thermal conductivities for: pure water, fully ionized matter, and constant electron conductivity.}
{Convective mixing complicates the relation between conductivity, evolution, and radius. For hot forming planets with a large composition gradient, where the heavy-element mass fraction changes gradually from the core to the envelope, convective mixing has a large impact on the radius evolution. In this case, the thermal conductivity is less relevant and the radii converge to similar values after billions of years. For cold forming planets or narrow composition gradients, convective mixing is less efficient. If the composition profile is not altered significantly, the thermal conductivity becomes critical. It determines how much energy can be trapped beneath stable composition gradients. For intermediate initial entropies, high thermal conductivity inhibits convection.}
{Further work is required to determine the thermal conductivity for various mixtures expected in sub-Neptune and Neptunes at high densities and temperatures. In addition, further constraints on the entropy and composition profile after formation can reduce the degeneracy of the planetary evolution, in particular, the dependence of the radius with time.}


\maketitle
\nolinenumbers

\section{Introduction}
Sub-Neptunes and Neptunes are among the most common planetary types in the current sample of over  $6000$ detected exoplanets. A  key objective of exoplanetary science is to understand their internal structure and origin by measuring their mass, radius, age, and atmospheric composition. However, determining their composition is challenging because multiple solutions can fit the same observational data. To reduce this degeneracy in bulk composition and internal structure, it is important to connect results form planet formation theories with planetary evolution simulations, and then link these to observations representing the planet's current-state. 

Exoplanet interiors are often modeled as separated layers of distinct composition. Each layer is assumed to be homogeneous in composition and fully convective, allowing the temperature profile to be approximated using the adiabatic gradient. However, several arguments challenge this layered structure.
For instance, the water distribution within planets is still under debate. Under high pressure water can preferentially dissolve into the core \citep{Lou+24} while also being produced at a magma hydrogen interface \citep{Horn+25}.
Additionally, advanced formation models indicate that composition gradients naturally arise from the accretion of solids and hydrogen-helium gas \citep[e.g.][]{Ormel+21, VallettaHelled22}.
Other processes taking place during the planetary evolution such as rain-out and demixing can also lead to the build up of boundary layers \citep[e.g.][]{Piaulet-Ghorayeb25, Arevalo+26}. Indeed, it was proposed that water rains out of the Hydrogen dominated envelope inside Uranus. This creates a sharp transition between a water rich and water poor layer \citep{CanoAmoros+24, Howard+25}. Such composition gradients can inhibit large scale convection\citep{StevensonSalpeter77, Guillot95, Markham+22}.
Recently, randomly generated density profiles of Uranus and Neptune that match the available gravity data show large regions without convection \citep{MorfHelled25}. This supports the claim that planets are not fully convective. 
Therefore, the temperature gradient is expected to deviate from the adiabatic gradient.

In non-convective regions, the energy transport depends on thermal conduction and radiation. In the outer envelope, where the density is small, most of the energy is carried by thermal photons that diffuse throughout the planet. This process is modeled using tables for the Rosseland mean opacity \citep{Ferguson+05, Freedman+14, Marigo+2024}.
At higher densities and temperatures, photon diffusion is inefficient and thermal conductivity dominates.
The thermal conductivity is often modeled by accounting for free electrons. However, also the atomic nuclei can contribute to the thermal conductivity, where the contribution depends on the exact planetary composition \citep{Ross+1984, Stamenkovic+2011, French2019}. For example, it was shown that neglecting the nuclei contribution in water, significantly underestimates the thermal conductivity \citep{French2019}. We summarize the contribution from the nuclei or lattice vibrations as ``vibrational conductivity".

In \citet{EberleinHelled25} \citep[erratum][]{EberleinHelled25Errata} (hereafter Paper I) we simulated the evolution of sub-Neptunes and Neptunes assuming that they have composition gradient that is stable against convection. For these types of planets, the composition gradient is expected to be relatively close to the atmosphere, i.e.,  above most of their internal energy. Hence, the treatment of the energy transport is crucial for their evolution. We compared four commonly used approaches to model the thermal conductivity. We found significant deviations in the thermal evolution of sub-Neptunes and Neptunes, which lead to radius difference of up to 20\% depending on the chosen model. Furthermore, we showed that a layer with low conductivity high up in the planet makes the initial entropy state in the deep interior more important. A hotter start can inflate the radius over a long period of time. However, the results presented in Paper I neglect the possibility of convective mixing and therefore the change of the internal structure with time. The objective of this study is to include this effect and then test the importance of the thermal conductivity in a self-consistent model when mixing is considered. 

Our paper is organized as follows.
In section \ref{sec:Methods} we describe the key aspects of the model. In section \ref{sec:Results} we present the results and address double diffusive convection in section \ref{sec:DDC}. We discuss our results in section \ref{sec:Discussion} and present our summary and conclusion in section \ref{sec:Summary}.

\section{Methods}\label{sec:Methods}
We followed the general procedure presented in Paper I. To model the evolution of sub-Neptunes and Neptunes we solved the stellar evolution equations \citep[e.g.]{Kippenhahn+2013} using the code Modules for Experiments in Stellar Astrophysics (MESA) \citep{Paxton+2011, Paxton+2013, Paxton+2015, Paxton+2018, Paxton+2019, Jermyn+2023}. For the H-He equation of state (EoS) we used the tables first presented in \citep{Mueller+2020a,Mueller+2020b} in the updated version \citep{MuellerHelled2021, MuellerHelled2024}, which includes non ideal interactions \citep{ChabrierDebras2021}. For the heavy-element EoS we assumed a fixed 50/50 water-to-rock ratio represented by H$_2$O and SiO$_2$ tables \citep{More+1988, Vazan+2013}. The tables were implemented in MESA based on a modified version of the extension \verb|custom EoS| \citep{KnierimHelled2024, Helled+25}.

For the atmosphere boundary condition we used the irradiated gray model \citet{GuillotHavel2011} with a ratio between the visible and thermal opacity of $\kappa\low{V}/\kappa\low{th}=0.03$ based on the fit provided by \citet{PoserRedmer2024}. We assumed an equilibrium temperature of $T\low{eq}=400$K.
A detail discussion on the model and its simplifications including the choice of our atmosphere, discrepancy between the assumed material for EoS and thermal conductivity, uncertainty in the initial conditions, choice of the EoS, and demixing, can be found in Paper I. 

\subsection{Thermal conductivity and opacity}
\begin{table}[hbt]

    \caption{Models used for the vibration and electron conductivity.}
    \begin{tabular}{cccc}
        \hline\vspace{1px}
        Conductivity & $k\low{vib}$ & $k\low{elec}$ & References\\\hline
        Cond-1 & H$_2$O  & partially ionized H$_2$O & (1), (2)\\
        Cond-2 & - & fully ionized & (3)\\
        Cond-3 & - & constant $k\low{elec}=4$ $\frac{\text{W}}{\text{mK}}$ & -\vspace{1px}\\
        \hline
    \end{tabular}
    \vspace{\baselineskip}
    \tablebib{
    (1)~\citet{FrenchRedmer2017},
    (2)~\citet{French2019},
    (3)~\citet{Cassisi+2007}, privately communicated by A.Y. Potekhin to the MESA developers. 
    }

    \tablefoot{
     AESOPUS2.1 is used for the radiative opacity. For models without a vibrational conductivity the summand $k\low{vib}$ is not considered when inferring the total conductivity.
    }

    \label{tab:ModelSetups}
\end{table}
\begin{figure}[hbt]
    \centering
    \includegraphics[width=\linewidth]{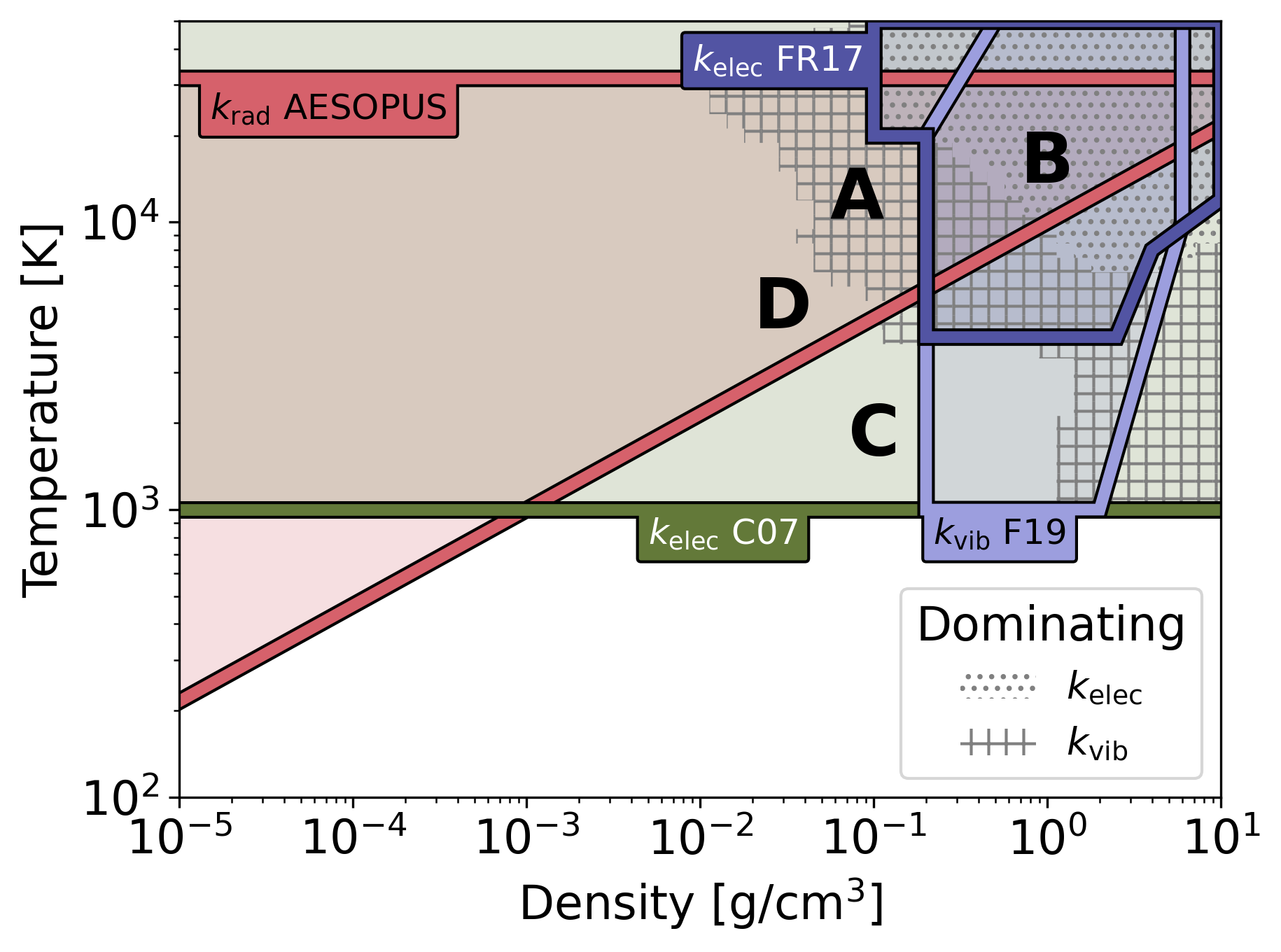}
    \caption{Boundaries of the different conductivity sources in density-temperature space. The red area ($k\low{rad}$ AESOPUS) shows the tabulated region from the AESOPUS2.1 tables \citep{Marigo+2024, EberleinHelled25}, the dark blue ($k\low{elec}$ FR17) and light blue ($k\low{vib}$ F19) area shows the region with ab initio data points for partially ionized water \citep{FrenchRedmer2017, French2019}. The green ($k\low{elec}$ C07) area shows the tabulated electron conductivity as implemented in MESA \citep{Cassisi+2007}. Hatched regions with squared lines ($k\low{vib}$ dominated) or dots ($k\low{elec}$ dominated) indicate where $k\low{vib}$ or $k\low{elec}$ contribute more than $50\%$ to the total conductivity assuming the Cond-1 model with $Z=20$ for the radiative conductivity. The density temperature regimes around A, B, C, and D are of particular interest and are further discussed in Section \ref{sec:Discussion}.}
    \label{fig:ConductivityRegionsSimple}
\end{figure}
\begin{figure}[hbt!]
    \centering
    \includegraphics[width=1\linewidth]{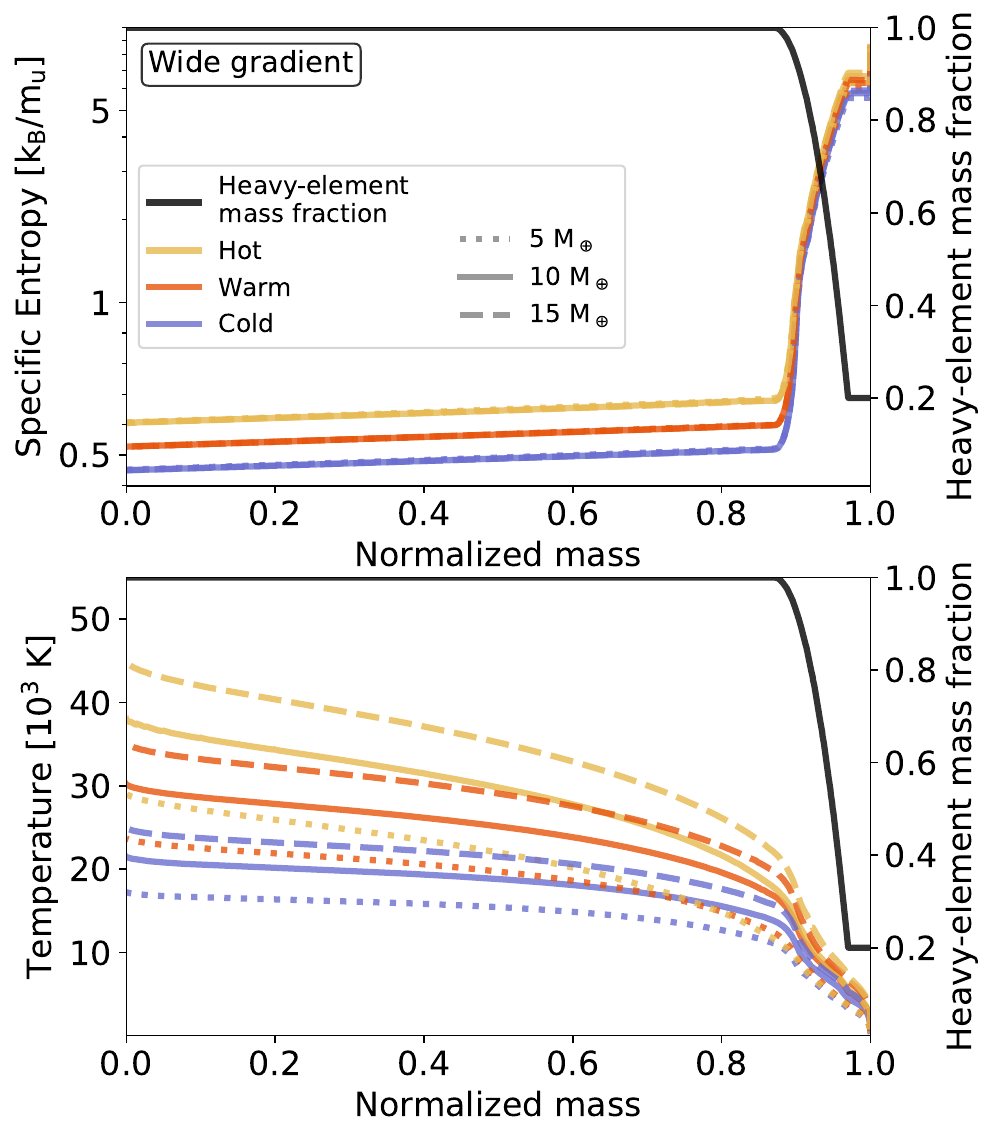}
    \caption{
    Initial profiles for the wide composition gradient, showing specific entropy (top) and temperature (bottom) as a function of normalized mass. The heavy-element mass fraction is overlaid in black in both panels, with its axis on the right. The colors blue (cold), orange (warm), and yellow (hot) correspond to different primordial entropies. The line styles dotted (5 M$_\oplus$), solid (15~M$_\oplus$), and dashed (10 M$_\oplus$) indicate different planetary masses.}
    \label{fig:InitialEntropyCompostionAndTemperatureProfiles}
\end{figure}

\begin{figure*}[t]
    \centering
    \includegraphics[width=1\linewidth]{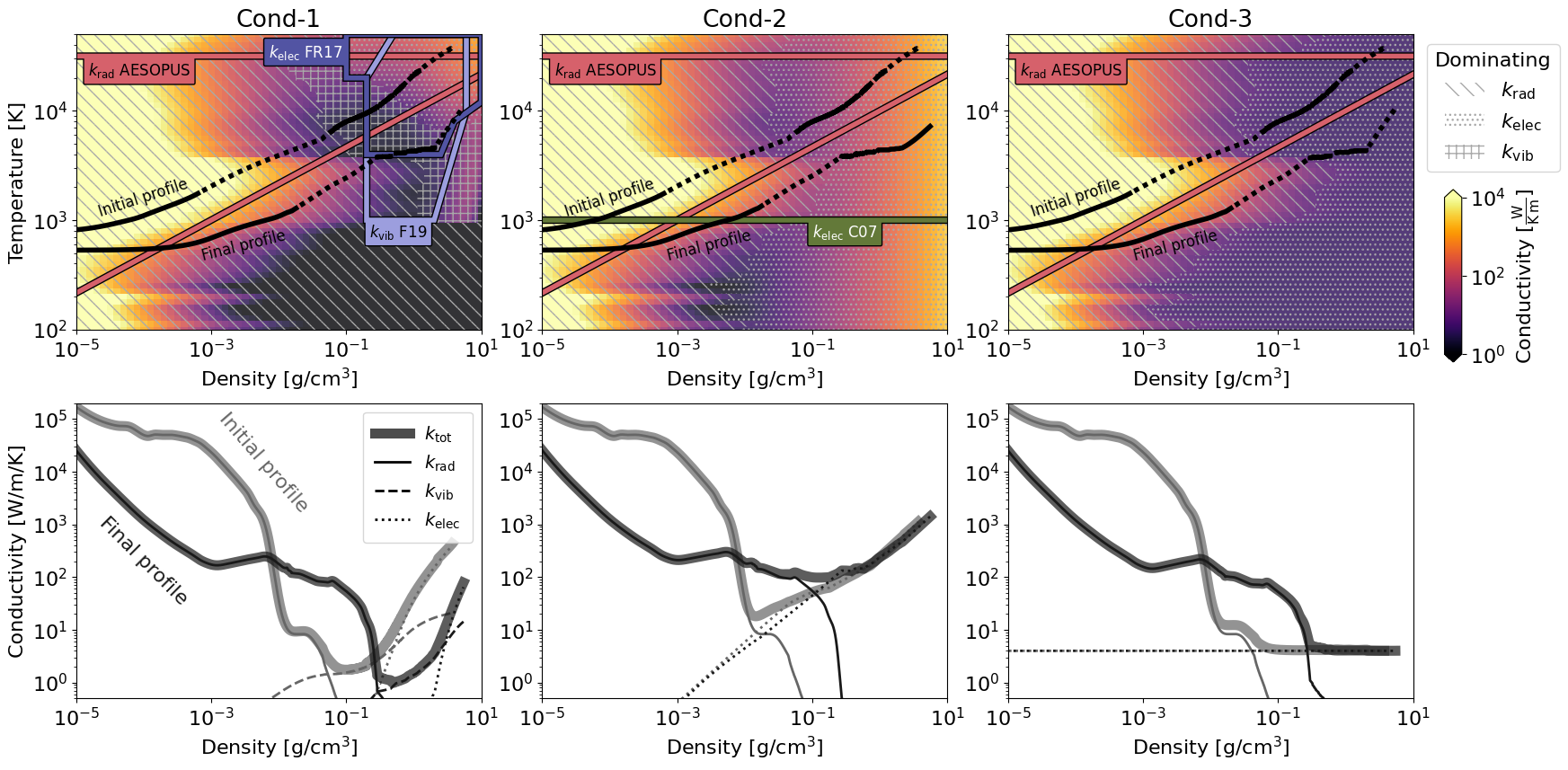}
    \caption{Thermal conductivity as a function of temperature and density (top row), and as a function of density along the initial and final planet profile (bottom row).\\
    {Top:} Total conductivity heatmap as a function of density and temperature for the three different conductivity models (from left to right): Cond-1, Cond-2, and Cond-3. The color represents the total conductivity value. Black lines show an example temperature-density profile for a planet with $M\low{p}=10\,\ME$ under the hot start scenario, calculated with the respective conductivity model: AESOPUS \citep{Marigo+2024, EberleinHelled25}, FR17 \citep{FrenchRedmer2017}, F19 \citep{French2019}, and C07\citep{Cassisi+2007}. The upper line (Initial profile) corresponds to the initial state of the evolution, while the lower line (Final profile) represents the state after 10$\,$Gyr. The dotted line indicates a convective region within the planet. Hatched regions with diagonal lines (radiation dominated), squared lines (vibration dominated), and dots (electron dominated) indicate where a specific conductivity mechanism contributes more than $50\%$ to the total conductivity. Colored lines with labels mark the boundaries of the conductivity models similar to Figure \ref{fig:ConductivityRegionsSimple}. 
    For this plot we assume $Z=0.20$ for the radiative opacity and $Z=1.0$ for the electron opacity of C07. We note that the the distribution of heavy-elements changes within the planetary interior.\\
    {Bottom:} Conductivity as a function of density for the initial (gray) and final (black) profiles shown in the top panel. The thick solid line shows the total conductivity along the density-temperature profile of the planet, while the thin solid (radiative), dashed (vibrational), and dotted (electron) lines show the individual conductivity contributions.}
    \label{fig:ConductivityRegions}
\end{figure*}

In non-convective regions the thermal transport depends on the total thermal conductivity $k\low{tot}$: 
\begin{equation}
    k\low{tot}=k\low{rad}+k\low{vib}+k\low{elec},\label{eq:k_tot}
\end{equation}
which is the sum of the conductivity contribution from photons $k\low{rad}$, vibrations in a dense fluid $k\low{vib}$ (lattice or nuclei), and electrons $k\low{elec}$. The radiative opacity is related to the radiative conductivity by: 
\begin{equation}
    \kappa\low{rad}=\frac{16\sigma}{3}\frac{T^3}{\rho k\low{rad}}, 
\end{equation}
with $\sigma$ being the Stefan-Boltzmann constant, $T$ the temperature, and $\rho$ the density. For the vibrational and electron conductivity, we used three different models. We summarize the three models in Table \ref{tab:ModelSetups} and the valid data region of each contribution in Figure \ref{fig:ConductivityRegionsSimple}.
For the radiative opacity $k\low{rad}$, we used tables calculated with the AESOPUS2.1 web interface \citep{MarigoAringer2009, Marigo+2024, EberleinHelled25}. In Appendix \ref{app:AesopusExtension} we describe how we treat high density values that fall outside the tabulated area. 
For the vibrational conductivity, we used an empirical fit to the ab initio calculated conductivities of water \citep[eq. 7 in ][]{French2019}. For the electron conductivity, we used three different models. First, the results for partially ionized water \citep[eq. 30 in][]{FrenchRedmer2017}. Second, the fully ionized electron conductivity implemented in MESA \citep[privately communicated by A.Y. Potekhin]{Cassisi+2007}. Third, a constant electron conductivity with a value that is expected for Earth's core-mantel boundary \citep[e.g.][]{Stevenson+1983, Lobanov+2021}. In Paper I, we had a fourth conductivity model, that included the vibrational conductivity model based on \citet{Stamenkovic+2011} for MgSiO$_3$. We concluded that this model is less relevant because the conductivity contribution from water is much higher and therefore more dominant. 
Additionally, the reference conductivity of MgSiO$_3$ is given at $\rho=3.87\,\text{g/cm}^3$ and $T=2000\,$K, which is at least $\sim1000\,$K below the temperatures reached in most of our models at similar densities.

\subsection{Convection}
We considered convective mixing and changed the composition in convective regions accordingly. A region is unstable against large scale convection if the Ledoux criterion is fulfilled: 
\begin{equation}
    \nabla\low{T}\geq\nabla\low{ad} + \frac{\varphi}{\delta}\nabla_\mu,\label{eq:Ledoux}
\end{equation}
with the temperature gradient $\nabla\low{T}=\frac{\dd\ln T}{\dd\ln P}$, the adiabatic temperature gradient $\nabla\low{ad}$, the mean-molecular weight gradient $\nabla_\mu=\left(\frac{\dd\ln \mu}{\dd\ln P}\right)$, and the thermodynamic derivatives $\delta=-\left(\frac{\partial\ln\rho}{\partial\ln T}\right)_{P,\mu}$ and $\varphi=\left(\frac{\partial\ln\rho}{\partial\ln \mu}\right)_{P,T}$ \citep{Ledoux47, Kippenhahn+2013}. Given a local luminosity $l$ and pressure $P$ the logarithmic temperature gradient by thermal conduction $\nabla\low{th}=\left(\frac{\dd\ln T}{\dd\ln P}\right)\low{th}$ is related to the conductivity by:
\begin{equation}
    \nabla\low{th}=\frac{1}{4\pi}\frac{lP}{\rho T k\low{tot}}.\label{eq:ThermalGradient}
\end{equation}
In stable region we set $\nabla\low{T}=\nabla\low{th}$. As a result, the possibility of  double diffusive convection (semi-convection) is not considered (see detailed discussion on double diffusive convection in section 4).
We assumed a mixing length of $\alpha\low{MLT}=0.1H_P$ where $H_P$ is the pressure scale-height. 
Simulating convective mixing in 1d hydrostatic codes is often challenging. We used the `extension `\verb|gentle_mixing|` to MESA \citep[][]{KnierimHelled2024, Helled+25} to limit the maximal change in the composition profile between two time steps. The extension controls the time-step and convective diffusion parameter to avoid sudden jumps in the composition that lead to convergence issues. We set the maximum squared difference of two consecutive composition profiles to $h^2_i=\int_0^M\left(X'_i-X_i\right)^2\dd m/M=0.001$, where $X_i$ and $X_i'$ are mass fractions of specie $i$ before and after a time step, respectively. Furthermore, we limit the maximum time step to $\Delta t=10^5$ years during the first $t=0.1$ billion years.

\subsection{Initial model}
We considered three different planetary masses of $M\low{p}=5, 10, 15\,\ME$. The envelope heavy-element mass fraction is set to $Z\low{env}=0.20$, and in the deep interior to $Z\low{int}=1.0$, where both regions are connected by a transition region with a smooth composition gradient. 
We considered three different composition gradients: (i) a wide transition that extends over a region with a mass of $\Delta q=0.1M\low{p}$, (ii) a medium transition $\Delta q=0.01M\low{p}$, and (iii) a narrow transition $\Delta q= 0.001M\low{p}$.  In all the models, the total heavy-element mass fraction in the planet is $Z\low{bulk}=0.95$.  The wide and narrow composition gradients are the same as in Paper I\footnote{We note that there was a typo in the text of Paper I regarding the width of the narrow model: the value that was used is $\Delta q=0.001$ and not $\Delta q=0.01$ as originally stated.}. We investigate the dependence of the evolution on the gradient width in further detail in Appendix \ref{app:GradientWidth}.

We used three different initial specific entropies, which we refer to as "cold", "warm", and "hot", respectively. Contrary to Paper I, we used an increasing entropy gradient towards the surface to aid numerical stability in the convective mixing phase. This leads to configuration of a stable planetary interior, with convective zones growing predominately inwards from the surface rather than emerging within the composition gradient. Before setting the composition profile we set the initial entropy $s\low{i}(q)$ as a function of the normalized mass coordinate $q=m/M\low{p}$ using the relation
\begin{equation}
    s\low{i}(q)=s\low{center,i}+q\Delta s,
\end{equation}
with the initial central entropy $s\low{center,i}$ and the slope set to $\Delta s=0.1\,\text{k}\low{B}/m\low{u}$. The parameter $s\low{center,i}$ was chosen such that $s\low{i}(q=0.8)=0.5,\,0.6,\,0.7\,\text{k}\low{B}/m\low{u}$. Using this approach the initial entropy slightly below the composition profile (at $q=0.8)$ is the same as in Paper I.
The initial models were created using the relax options of MESA in multiple steps (see Paper I for further details). 
Figure~\ref{fig:InitialEntropyCompostionAndTemperatureProfiles} shows the initial entropy and temperature profile for the wide composition gradient models.

\section{Results}\label{sec:Results}
\subsection{Comparison of conductivity models}\label{sec:ConductivityRegions}
Figure \ref{fig:ConductivityRegions} shows the total conductivity using each conductivity model. To illustrate which density temperature regions are important for the evolution of planets we overlay example profiles at the beginning of the simulation and after 10 billion years. We note that these models are for illustration and vary for different input parameters and assumptions. For example, much colder planets or faster cooling planets will enter regions with lower temperatures.
The colored lines indicate the valid regions of the models, outside these regions the models are extended.

For Cond-1, the vibrational conductivity divides the region that is dominated by radiation and electron conduction. Given its extensive coverage in the density-temperature space, it can not be neglected. Therefore, a planet that contains significant amounts of water should have a large region where non-convective energy transport is dominated by the vibrational conductivity. However, the fits of the vibrational conductivity are extended beyond the recommended range by the authors (see $\rho\lesssim0.2$g/cm$^3$ and $T\gtrsim4\,000$K, and $\rho\gtrsim2$g/cm$^3$ and $T\gtrsim1\,000$K). In these extrapolated regions no data points validate the fit, yet the predicted values exceed those of the radiative conductivity of the AESOPUS2.1 tables. Electron conductivity dominates only at high densities and high temperatures.
A significant part of the planet is outside the region that is covered by any of the tabulated data. As the planet cools, this region becomes larger.

For Cond-2, the electron conductivity dominates a larger part of the temperature-density space. In particular, the vibrational conductivity of Cond-1 is insignificant compared to the electron conductivity of Cond-2. The electron conductivity of \citet{Cassisi+2007} is tabulated for $T\geq10^{3}$\,K. Hence in the case of Cond-2, the part of the planet around $\rho\sim10^{-3}\,\text{g/cm}^3$ reaches temperature-density values where neither the electron conductivity nor the radiative conductivity is valid. 

In the case of Cond-3, the electron conductivity dominates at higher densities, where radiation transport is ineffective. Again, with this approach, a large part of the planet is modeled outside the original AESOPUS2.1 table, and conductivities must be extrapolated.

\subsection{Convective mixing}
\begin{figure}[hbt]
    \includegraphics[width=0.97\linewidth]{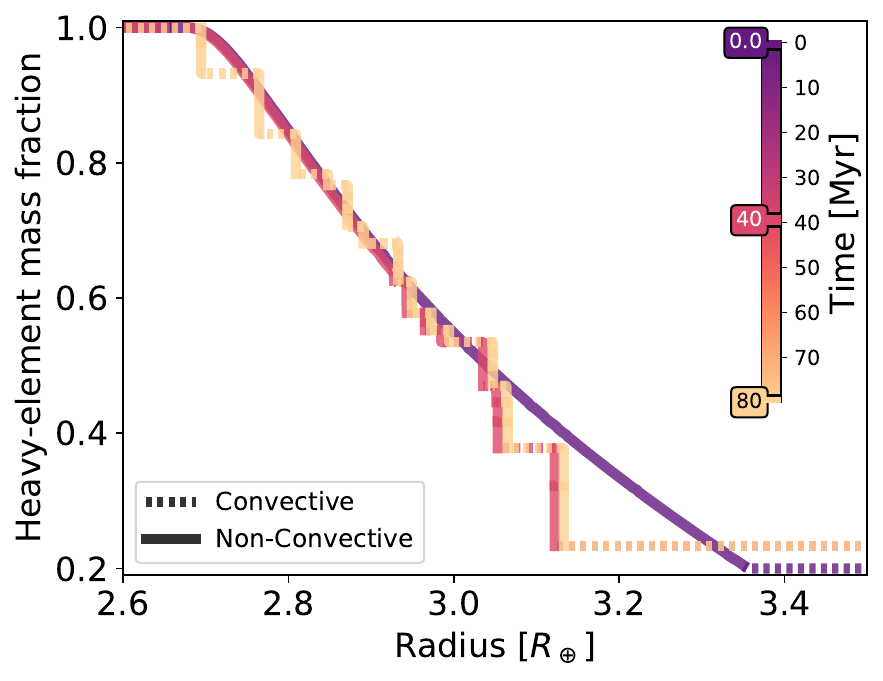}
    \includegraphics[width=1\linewidth]{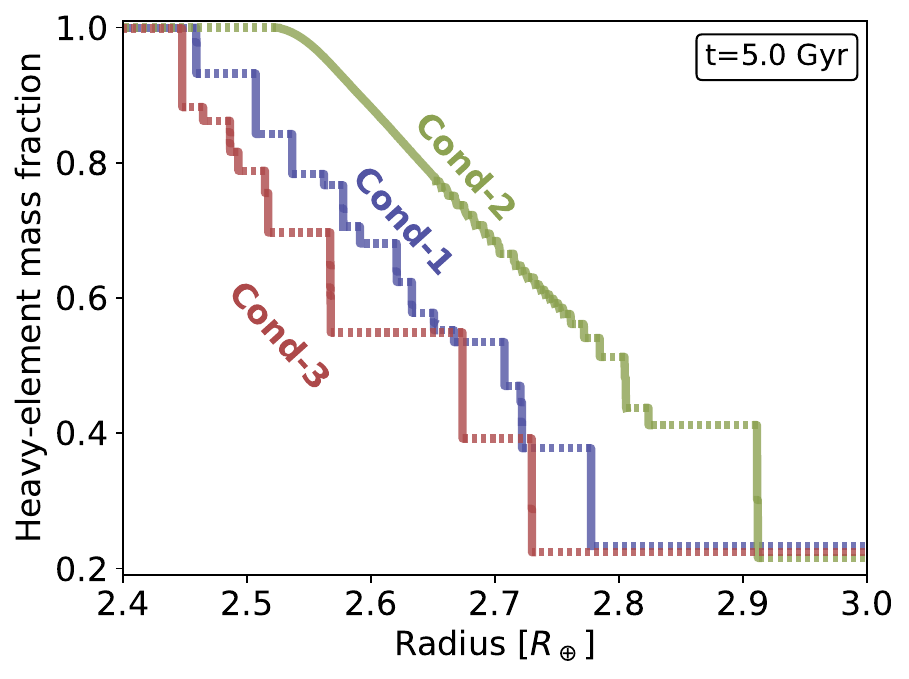}
    \caption{ 
    Heavy-element mass fraction vs.~radius at different times for the Cond-1 model (top) and for the three different models at $t=5\,$Gyr (bottom). Solid lines indicate non-convective regions, while dashed lines indicate convective regions. This plot corresponds to the simulations of a planet with $M\low{p}=10\,\ME$, a wide composition gradient and a warm start.
    }
    \label{fig:ConvectiveStaircases}
\end{figure}
The key difference from Paper I is the consideration of convective mixing. As the planet cools, the convective region moves from outside in and erodes the composition gradient. This increases the heavy-element mass fraction in the envelope and can create convective staircases \citep[e.g.][]{Vazan+18, Mueller+2020a, KnierimHelled2024, Arevalo+25a}. These staircases should not be confused with the process of double-diffusive convection \citep{Garaud18}, which is not included in our simulations. 
The upper panel of Figure \ref{fig:ConvectiveStaircases} shows the composition profile at different times. Convective zones appear in the region that was previously a smooth composition gradient. Small non-convective regions separate the convective zones with jumps in the heavy-element mass fraction. We find that this process is very sensitive to the used model assumptions. Nevertheless, we can identify certain trends. If the planet starts hotter, fewer and larger steps appear during the planetary  evolution. If the planet starts very cold, some of the composition gradient remains with a large non-convective region.

Furthermore, the thermal conductivity influences when a region becomes unstable against convection. To illustrate this, we show  the final composition profile using the different conductivity models in the lower panel of Figure \ref{fig:ConvectiveStaircases}. Because Cond-2 has a higher conductivity, the temperature gradient in non-convective regions is shallower (see eq. \ref{eq:ThermalGradient}). It follows that the shallow temperature gradient in non-convective regions leads to higher stability against convection (see equation \ref{eq:Ledoux}). Therefore, the composition profile does not develop a convective staircase throughout the entire composition gradient. The lower conductivities of Cond-1 and Cond-3 lead to steeper temperature gradients that fulfill equation \ref{eq:Ledoux} such that staircases appear. 

\subsection{Radius evolution}
\begin{figure}[hbt]
    \centering
    \includegraphics[width=1\linewidth]{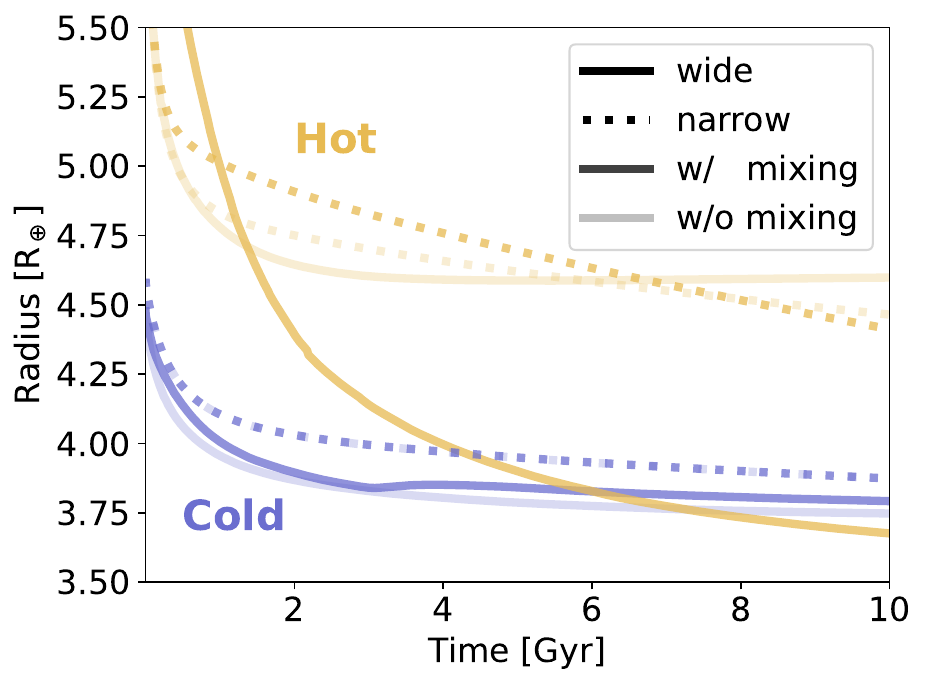}
    \caption{ 
    Radius over time with and without mixing. The simulations with mixing are shown in dark colors while the simulations without mixing are shown in light colors. The solid lines represent the models with a wide composition gradient while the dotted lines represent the model with a narrow composition gradient. The blue color indicates a cold start, while the yellow color indicates a hot start. The simulations for this plot assume $M\low{p}=10$M$_\oplus$, a wide gradient, and Cond-1.
    }
    \label{fig:RadiusMixingCompariosn}
\end{figure}

Next, we compare the radius evolution with and without mixing.  Figure \ref{fig:RadiusMixingCompariosn} shows the radius evolution for different initial entropies and a wide and a narrow composition gradient.
For a cold start the entropy is not high enough to erode the composition gradient. In this case the composition profile remains mostly intact and therefore the difference between the simulations with and without mixing are low. 
The narrow gradient is too steep to be eroded  by convection and a convective staircase does not appear. Hence a thermal boundary layer remains between the envelope and the deep interior such that the radius evolves similar to the case without mixing.
For the hot start with a wide gradient the entropy is high enough to erode some of the composition gradient and to create a composition staircase. This significantly increases the thermal transport such that the interior can effectively cool down. The radius contracts over the entire evolution.

\begin{figure}[hbt!]
    \centering
    \includegraphics[width=1\linewidth]{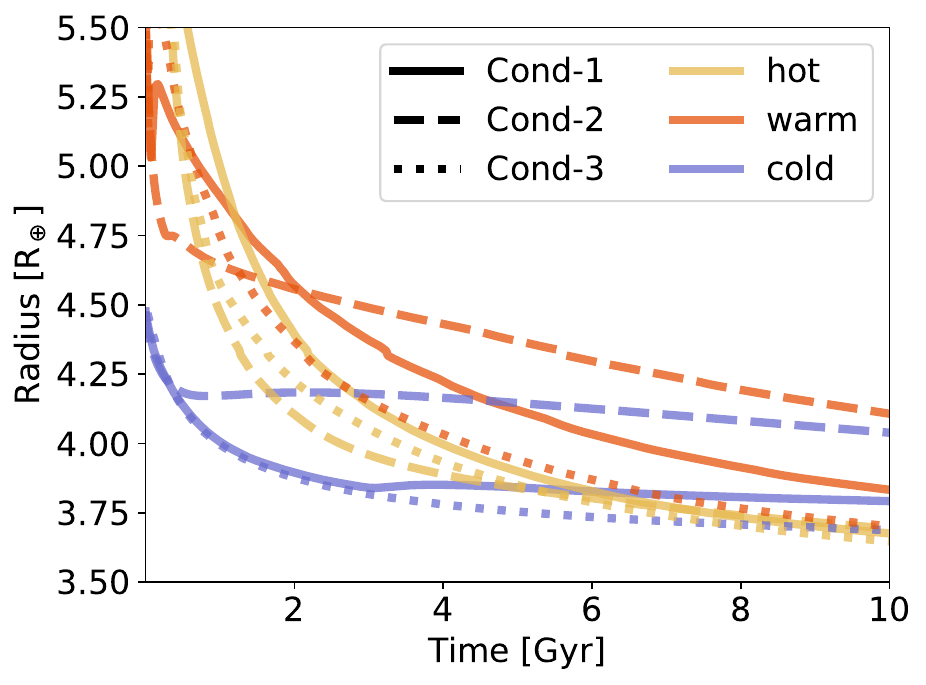}
    \caption{Radius evolution for a planet with $M\low{p}=10\,\ME$ and a wide gradient. The blue, orange, and yellow lines represent the cold, warm, and hot initial entropies, respectively. Solid lines correspond to the Cond-1 model, dashed lines to the Cond-2 model, and the dotted lines to the Cond-3 model.}
    \label{fig:RadiusEvolutionM10}
\end{figure}

In Figure \ref{fig:RadiusEvolutionM10} we show the radius evolution for a planet with $M\low{p}=10\,\ME$ including convective mixing (see Appendix \ref{app:RadiusEvolutionM05M15} for $M\low{p}=5\,\ME$ and $M\low{p}=15\,\ME$). As seen before if the planet starts cold the radius evolution is similar as in Paper I. Trapped heat creates smaller radii for Cond-1 and Cond-3 compared to Cond-2. The deviation from the previous paper arise at higher initial entropies. 
In case of the warm scenario, the Cond-2 model has a short phase where the radius increases at $t\approx0.5\,$Gyr. This increase happens when the core becomes convective and the deep interior can rapidly loose energy. We find this behavior in most of the warm and hot initial models. In the case of the warm Cond-3 model, the low conductivity gives rise to convective instabilities. The resulting composition profile contains few steps with large convective zones (see Figure \ref{fig:FinalCompositionProfiles} in the appendix). This results in an efficient energy transport through the staircase such that the radius evolution is determined by the cooling through the atmosphere.
The hot models start much more inflated but cool down over the lifetime of the planet and eventually evolve to similar radii. Within the first $\sim100\,$Myr, the mixing process is complete and all hot models develop convective staircases (see Figure \ref{fig:FinalCompositionProfiles} in the appendix). The small non-convective regions between these staircases are not large enough to effectively trap heat. The radius evolution is therefore mostly governed by the cooling through the atmosphere. This explains why the different conductivity models result in a similar radius evolution.

\begin{figure}
    \centering
    \includegraphics[width=1\linewidth]{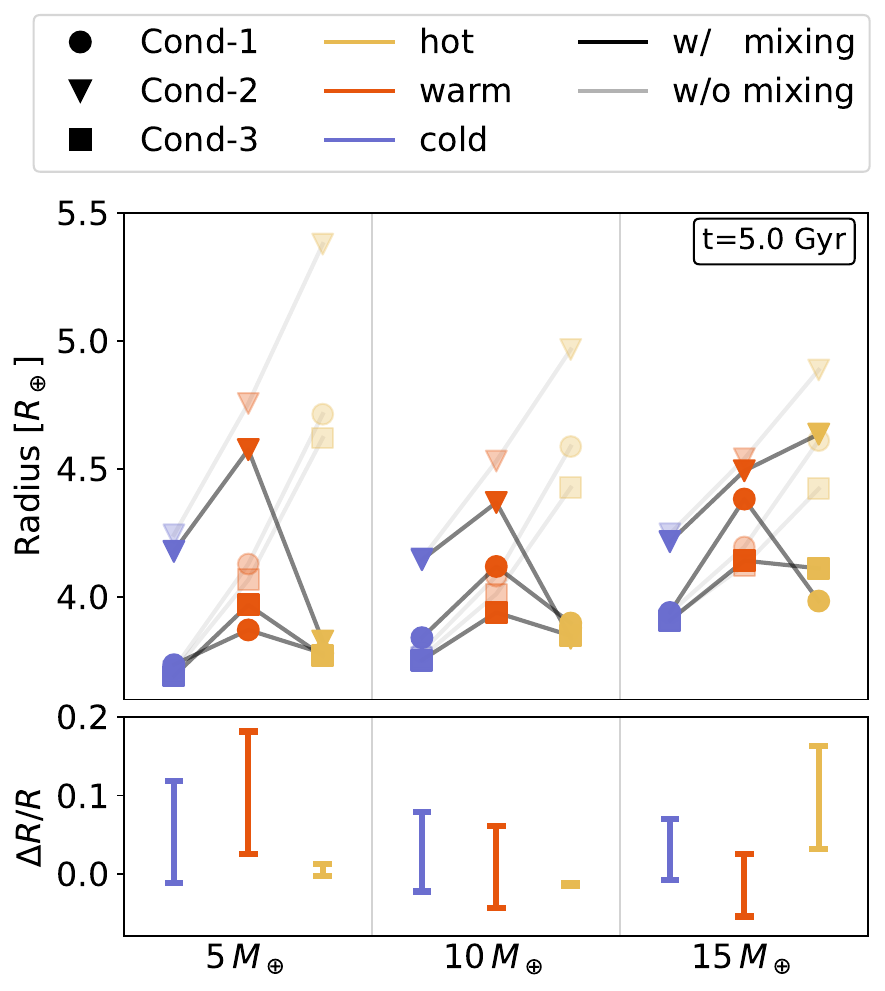}
    \caption{Planetary radius at $t=5$ Gyr (top) and relative difference between the conductivity models (bottom). The left ($M\low{p}=5\,\ME$), middle ($M\low{p}=10\,\ME$), and right ($M\low{p}=15\,\ME$) columns show different planet masses. The markers circle (Cond-1), triangle (Cond-2), and square (Cond-3) represent the conductivity model. The light shades in the upper panel show results without convective mixing. We connect the different entropies with the same models for visibility reasons. The relative difference in the radius is calculated using $\Delta R/R = (R' - R\low{Cond-1})/R\low{Cond-1}$, where $R\low{Cond-1}$ is the radius for Cond-1 and $R'$ the radius for Cond-2 or Cond-3.}
    \label{fig:RadiusSummary}
\end{figure}

Figure \ref{fig:RadiusSummary} shows the planetary radius at $t=5\,\text{Gyr}$ for different masses and conductivities, both with and without convective mixing. The results demonstrate that for high initial entropies, the thermal conductivity is less relevant for the late radius evolution (after a few Gyrs). 
This is because convective mixing can transform the composition gradient into staircases, where heat trapping is less efficient. 
Interestingly, for a $M\low{p}=15\,\ME$ planet, the hot Cond-2 model is stable against convection, and therefore the planet preserves most of its primordial composition profile (see Figure \ref{fig:FinalCompositionProfiles} in the appendix). 
The lower panel of Figure \ref{fig:RadiusMixingCompariosn} shows the  relative radius difference (see caption for details). 
For low initial entropies, most of the composition gradient is sustained, leading to a significant radius difference between the different conductivity models. Overall, the figure indicates that the minimal primordial entropy required to destabilize most of the composition gradient depends on the planetary mass.  

\section{Double diffusive convection}\label{sec:DDC}
In this work, we used the Ledoux criterion to distinguish between stable and convective regions. In stable regions, we assumed that energy transport occurs only by thermal diffusion and neglected material diffusion. In reality, regions that have been identified as stable could develop double diffusive convection (DDC) under some circumstances.  This can occur in regions where the thermal gradient is destabilizing, while the compositional gradient is stabilizing. The regimes that could develop DDC can be identified from the inverse density parameter $R_\rho^{-1}$ as \citep[e.g.][]{LeconteChabrier2012}:
\begin{equation}
    R_\rho^{-1}=\frac{\varphi}{\delta}\frac{\nabla_\mu}{\nabla\low{T}-\nabla\low{ad}}.
\end{equation} 
The transition between the purely diffusive energy transport regime and the DDC regime (layered or oscillatory) can be estimated using the critical inverse density parameter $R_{\rho,\text{crit}}^{-1}$
\begin{equation}
    R_{\rho,\text{crit}}^{-1}=\frac{Pr + 1}{Pr + \tau},
\end{equation}
where $Pr$ is the Prandtl number and $\tau$ is the ratio between solute to thermal diffusivity. 
Convection occurs when $R_\rho^{-1}\leq1$ (equal to Ledoux unstable), DDC occurs when $1<R_\rho^{-1}\leq R_{\rho,\text{crit}}^{-1}$, and stable when  $R_{\rho,\text{crit}}^{-1}<R_\rho^{-1}$ 
\citep[see][for a detailed discussion]{Rosenblum+11, Mirouh+12, LeconteChabrier2012}. 
The value of $R_{\rho,\text{crit}}^{-1}$ is very uncertain because it depends on the material properties that determine the Prandtl number and diffusivities. 

\begin{figure}
    \centering
    \includegraphics[width=\linewidth]{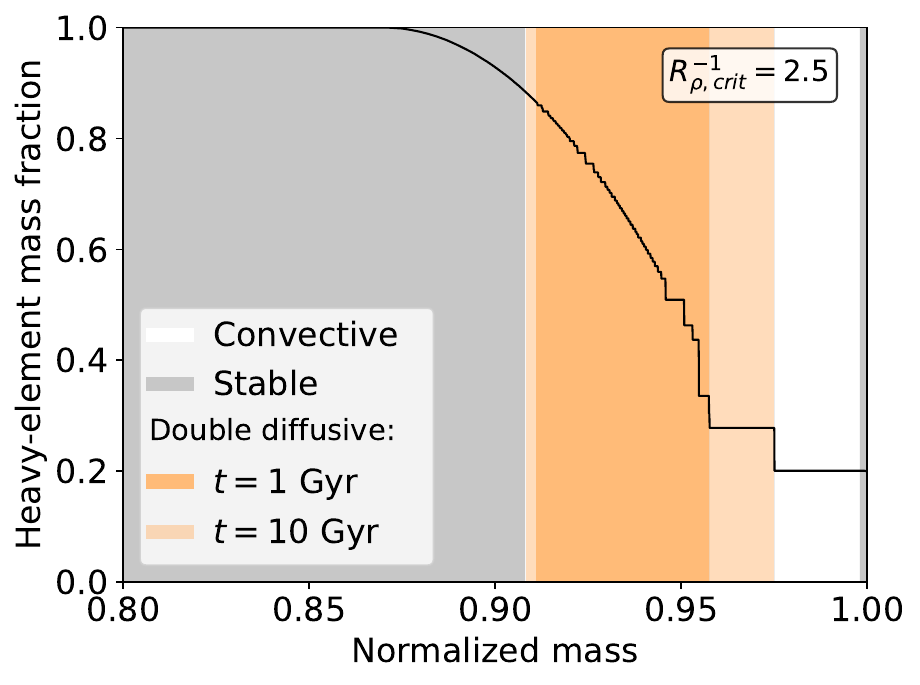}
    \caption{Heavy-element mass fraction as a function of normalized mass for the cold $M\low{p}=10\,\ME$ planet with the wide composition gradient at $t=1\,\text{Gyr}$ (dark orange) and $t=10\,\text{Gyr}$ (light orange). The background color white (convective), gray (stable) and orange (double diffusive) indicate the stability regimes assuming $R_{\rho,\text{crit}}^{-1}=2.5$.}
    \label{fig:Semiconvection}
\end{figure}

In the context of this work, the main difference in the radius evolution originates from whether a stable layer can be sustained over timescales of several 10$^9$  years. We find that hot models develop convective stair cases while colder models retain a large stable region, and therefore the possibility of DDC may be more relevant for these cold cases. We therefore explore how DDC could affect our results for the cold Cond-1 model with $M\low{P}=10\,\ME$ and the wide composition gradient. 
Figure \ref{fig:Semiconvection} presents the composition profile at two different ages of  $t=1\,\text{Gyr}$  (dark orange) and $t=10\,\text{Gyr}$ (light orange) indicating the regions where DDC could occur. We identify  the lower and upper bound of the region that can potentially become DDC unstable. For simplicity,  we adopted a constant value $R_{\rho,\text{crit}}^{-1}=2.5$ as suggested for Uranus using the same conductivity models for $k\low{vib}$ and $k\low{elec}$ \citep{FrenchNettelmann19}. However, we note that the value of $R_{\rho,\text{crit}}^{-1}$ is rather uncertain and would have a different value for different compositions, and also that its value is expected to vary with temperature and density. For dense water, values range from 1~$-$~7 \citep{FrenchNettelmann19}. 
We also note that the time when DDC could start within the planetary depends on the value of $R_{\rho,\text{crit}}^{-1}$. Higher (lower) values than considered here would lead to DDC occurring earlier (later).  However, we find that varying $R_{\rho,\text{crit}}^{-1}$ by a factor of 2 and 0.5 nearly does not change the size of the zone that could become DDC unstable, and therefore will not affect our conclusion (see Appendix \ref{app:SemiConvZonesExtended}). In addition, we note that the self-consistent implementation of DDC in planetary evolution models is non-trivial and is still being investigated \citep{LeconteChabrier2012, Wood+13, KurokawaInutsuka15, Fuentes+22, Anders+22, Tulekeyev+24, DudeHansen25, Fuentes25}. 

In DDC regions, both the thermal and compositional fluxes are higher relative to a fully stable configuration \citep{StevensonSalpeter77, Rosenblum+11, Mirouh+12, Garaud18}. In our particular example, we expect the thermal flux to be limited by the stable region. As a result, the radius evolution should be similar with and without DDC. However, we note that layered DDC (occurs at $R_{\rho}^{-1}$ close to 1) can lead to large scale convection. The exact details of DDC and its effect on planetary evolution are complex and the uncertainties and the effect on the evolution originating from the initial composition and entropy/temperature profile are likely more important. Again, we note that DDC is unlikely to change the results for the warm and hot models since they are mostly convective. The fact that we consider different conductivities highlights the importance of understanding heat (and material) transport when composition gradients exist, and we therefore encourage more studies that constrain $R_{\rho,\text{crit}}^{-1}$ for different compositions. We also hope that future work develops a more comprehensive implementation of DDC at planetary conditions.

\section{Discussion}\label{sec:Discussion}
\subsection{Thermal conductivity}
With the Cond-1 model, we tested the case where water dominates the non-convective thermal transport. In this case, there is a region within the planet where the thermal conductivity is governed by the vibration of the nuclei rather than by radiation or electron conduction. We identify a region with $\rho\leq0.2\,\text{g/cm}^3$ (Figure \ref{fig:ConductivityRegionsSimple}, region A) where the fit of \citet{French2019} for the vibrational conductivity becomes the dominant contribution to the thermal conductivity. Since these densities lie outside the fitted range, the extrapolated conductivity in this region is highly uncertain. It remains unknown which values of the vibrational conductivity are realistic and whether the radiative conductivity is smaller. The electron conductivity contribution from water always dominates the deep interior (Figure \ref{fig:ConductivityRegionsSimple}, region B). Because the thermal conductivity of pure water is likely a lower boundary of the real conductivity inside sub-Neptunes and Neptunes \citep{Scheibe+21}, it remains unclear how the conductivity would change if a mixture (e.g., rock+water) is considered.

The electron conductivity from \citet{Cassisi+2007} covers most of the required temperature-density regime, but its assumption of fully ionized material likely overestimates the conductivity. For example, at Earth's core-mantle boundary, the expected thermal conductivity ranges between $k\low{tot}=4-11\,\text{W/m/K}$ with vibrational contribution dominating at $k\low{vib}=3-10\,\text{W/m/K}$ \citep{Lobanov+2021}. This is in contrast with the Cond-2 model, where electron conductivity consistently exceeds vibrational contribution.

Pure water conductivity is a simplification and a lower limit \citep{Scheibe+21}, while fully ionized electron conductivity likely overestimates the true value. And, there is evidence suggesting that vibrational (or nuclear) contributions are non-negligible. We conclude that thermal conductivity under planetary interior conditions requires further study, in particular for mixtures expected in sub-Neptunes and Neptunes.

\subsection{Rosseland-mean opacity tables}
In Section \ref{sec:ConductivityRegions} we identified the  temperature-density regimes that are important for the planetary evolution. We identified a large portion within the planet where thermal transport is dominated by photons, but the radiative opacities do not extent to high enough densities (Figure \ref{fig:ConductivityRegionsSimple}, region C).
Although the opacities from \citet{Freedman+14} as implemented in MESA provide higher densities ($\log \tilde{R}\,[\text{g/cm}^3]\leq9$), they do not properly cover temperatures above $\log T\,[\text{K}]\geq3.6$. Additionally, Figure \ref{fig:OpacityDensityExtrapolation} shows that the high density regions have been extrapolated. 
The temperature density profile of a planet roughly follows curves with constant density parameter $\tilde{R}$. Therefore, it is a good indicator, where opacity calculations are required. All the planetary models presented in this study have values within $3\leq\log \tilde{R}\,[\text{g/cm}^3]\leq8$. We note that $\tilde{R}$ can have  much lower values in the planet's atmosphere. Higher values are expected for much colder and denser planets.
To properly account for the radiative transport of energy inside sub-Neptunes and Neptunes, Rosseland-mean opacity tables are required in the density parameter range at least up to $\log \tilde{R}\,[\text{g/cm}^3]\leq8$ for various heavy-element mass fractions. 

From Figure \ref{fig:ConductivityRegions}, we further constrain the temperature-regime where opacity calculations are required. At higher densities of $\rho\gtrsim0.1\text{g/cm}^3$, the temperature regime in which radiative energy transport dominates becomes smaller. In our setup with AESOPUS2.1, and the conductivity models, the highest required temperature is around $\sim5\,000\,$K. However, at lower densities of $\rho\lesssim0.1\text{g/cm}^3$ (Figure \ref{fig:ConductivityRegionsSimple}, region D), the regime in which radiative energy transport dominates extends to temperatures above $5000\,$K.

Furthermore, we note that the interior structure of sub-Neptunes and Neptunes is still relatively uncertain. It is possible that such planets have more complex interiors that include composition gradients and layers of different heavy-element mass fraction \citep{VallettaHelled22, CanoAmoros+24, MorfHelled25}. Therefore,  it is desirable to have opacity tables for various heavy-element mass fractions reaching values up to $Z=1$.

\subsection{Importance for exoplanets}
Connecting the radius, mass, and age (time) to the planetary internal structure is very important for connecting planet formation with current-day observations. Our study clearly shows that the radius evolution of sub-Neptunes and Neptunes strongly depends on the assumed conductivity and the initial entropy profile. In particular, for sub-Neptunes and Neptunes, the conductivity plays a key role.  In the case of intermediate-mass planets, the primordial composition gradient is expected to exist above a significant fraction of the mass compared to gas giant planets \citep[][]{HelledStevenson17, Ormel+21, VallettaHelled22}. Therefore, the gradient that acts as a thermal boundary layer is located above a significant amount of the internal energy budget. As a result, differences in the thermal transport (determined by the conductivity) can significantly affect the cooling. In addition, the density of the hydrogen-helium atmosphere is very sensitive to temperature. Hence, the hydrogen helium envelope strongly correlates the thermal flux from the deep interior with the radius. Therefore, we expect that the evolution and internal structures of intermediate-mass planets are particularly sensitive to thermal conductivities.

\section{Summary and conclusions}\label{sec:Summary}
We studied the impact of the thermal conductivity on the evolution of sub-Neptunes and Neptunes. We improve on previous results from Paper I by including the effect of convective mixing. The main findings of this paper can be summarized by:
\begin{itemize}
    \item The available data of thermal conductivity and radiative opacities are still insufficient for sub-Neptunes and Neptunes. Public available Rosseland-mean opacities do not cover high enough temperatures, high enough densities, or high metallicities. The thermal conductivity of mixtures (water - rock, water - hydrogen/helium, water - methane) is uncertain.
    \item The thermal conductivity affects convective mixing and the final composition profile. High conductivity can inhibit convection.
    \item The inferred effect on the radius is more complex than in Paper I. We identified two cases: First, for low entropies convective mixing does not create a convective staircase over the entire composition gradient. A low conductivity creates a thermal boundary layer (see also \citet{Scheibe+21} and Paper I). Second, for high entropies a convective staircase replaces the composition gradient. In this case, the energy transport into the envelope is enhanced. Depending on the shape of the final composition profile, the radius can converge to similar values regardless of the conductivity model.
    \item The composition gradient can be too step to be eroded by convective mixing. In this case a thermal boundary layer remains, and the thermal evolution of the planet strongly depends on the assumed thermal conductivity. 
\end{itemize}
Our results clearly indicate that further work is needed to better model the thermal transport inside sub-Neptunes and Neptunes. Improved calculations (and experiments) of the radiative opacities at higher densities and temperature for high metallicities are required. The determination of the thermal conductivity of various mixtures is also desirable. Finally, we note that in order to connect observations of ``evolved planets" using evolution simulations, further constrains on the initial entropy and compositions are needed.

\begin{acknowledgements}
This work was supported by the Swiss National Science Foundation (SNSF) through a grant provided as a part of project number 215634: \url{https://data.snf.ch/grants/grant/215634}.
We thank the referee for their helpful comments. 
We also thank Simon Müller and Henrik Knierim for many valuable discussions and technical support.\\
\textit{Software:} \verb|gentle_mixing| \citep{KnierimHelled2024, Helled+25},  MESA \citep{Paxton+2011, Paxton+2013, Paxton+2015, Paxton+2018, Paxton+2019, Jermyn+2023}, PyMesaReader, AESOPUS2.1 \citep{MarigoAringer2009, Marigo+2024}, Jupyter Notebook \citep{Kluyver2016, Granger+2021}, NumPy \citep{Harris2020}, Matplotlib \citep{Hunter2007}, Astropy \citep{Astropy2013, Astropy2018, Astropy2022}
\end{acknowledgements}

\bibliographystyle{aa}
\bibliography{bibliography}

\begin{appendix}
\onecolumn
\section{Extension of the radiative opacity tables}\label{app:AesopusExtension}
In Paper I we created a set of Rosseland mean opacity tables for planetary evolution simulations, using the AESOPUS2.1 code \cite{MarigoAringer2009, Marigo+2024}. The tables span a wide range of temperature with log$(T/$K$)=\left[2,\,4.5\right]$ and density parameter $\tilde{R}=\rho/(10^{-6}\,T/$K$)^3$ with $\log(\tilde{R} /($g/cm$^{3}))=\left[-8,\, 6\right]$. Planets usually reach higher values of $\tilde{R}$ during their evolution. Here we compare two possible extensions of the tables to values of $\log(\tilde{R} /($g/cm$^{3}))=9$. The first method assumes a constant opacity for values above the table limit such that $\kappa(T, \tilde{R} \geq 10^6 $g/cm$^3)=\kappa(T, \tilde{R}=10^6$g/cm$^3)$ (MESA default). The second method extrapolates linearly from the last two table entries in a logarithmic space with
\begin{equation}\label{eq:OpacityExtrapolations}
    \log \kappa(\log \tilde{R} \geq 6)=\log \kappa(6) + \frac{\log \kappa(6) - \log \kappa(5.8)}{0.2} \left(\log \tilde{R} - 6\right),
\end{equation}
where the dependence of $\kappa(\log \tilde{R})$ on $\log T$ is omitted for readability reasons and $\tilde{R}$ is measured in g/cm$^3$. A linear extrapolation of $\kappa$ instead of $\log\kappa$ would result in negative opacities.
\begin{figure*}[hbt!]
    \centering
    \includegraphics[width=1\linewidth]{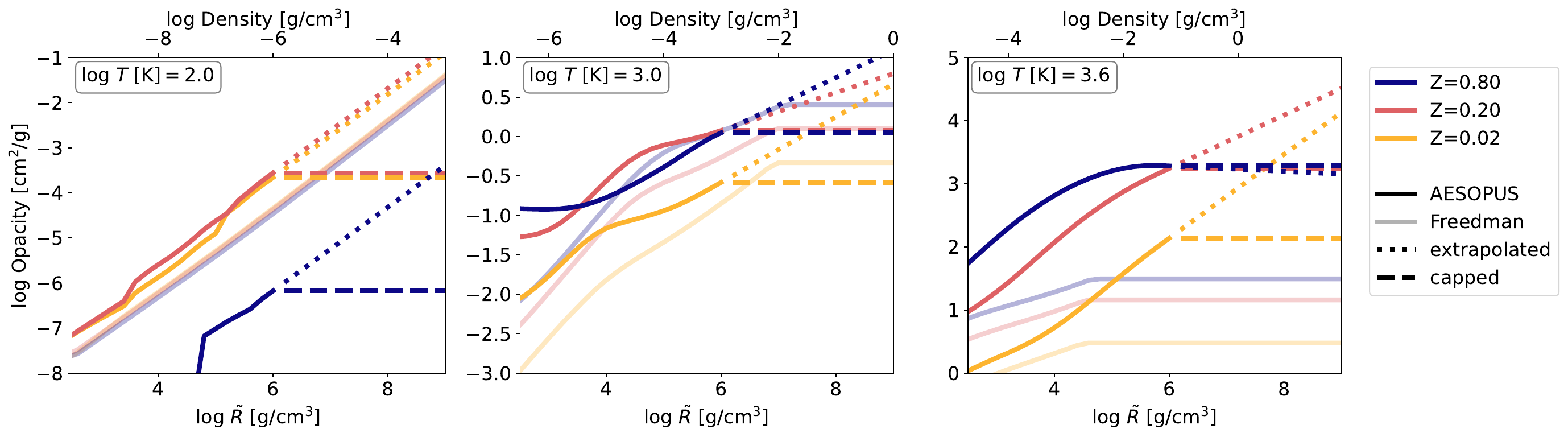}
    \caption{Radiative opacity as a function of the density parameter $\tilde{R}$ (with density shown on the upper x-axis) from the AESOPUS2.1 tables. For comparison, the Freedman opacity curves as implemented in MESA are shown in light shades. The three panels, from left to right, correspond to temperatures of $\log T $ [K]$=2,\, 3,$ and $3.6$. The blue, red, and yellow lines represent metallicities of $Z=0.80$, $0.20$, and $0.02$, respectively. Solid lines indicate regions with tabulated data points ($\log(\tilde{R} /($g/cm$^{3}))\leq6$), dotted lines show linear extrapolation on logarithmic values (see eq. \ref{eq:OpacityExtrapolations}), and dashed lines denote constant values outside the data region.}
    \label{fig:OpacityDensityExtrapolation}
\end{figure*}
In Figure \ref{fig:OpacityDensityExtrapolation} we show the scaling of the opacity with respect to the density. To compare our method of choice to extent the tables to higher densities we plot both methods in the region above $\log(\tilde{R} /($g/cm$^{3}))=6$. From these plots we conclude that the extrapolation seems to be the better choice. Keeping the opacities constant at high densities seems to significantly underestimate the opacities at high density. Especially at low temperatures, where the Freedman opacities show an increase in the opacity over multiple orders of magnitude. The Freedman opacities extend to $\log(\tilde{R} /($g/cm$^{3}))=9$ yet the tables have been extended by keeping the opacity constant at high densities. The density at which the opacity is constant is between $\log \rho [$g/cm$^3]=\left(-2.75,\,-1.5\right)$ depending on the temperature.

We note that the higher density opacities are less relevant where the thermal transport is not dominated by photons. This can either be the case when electron or vibrational thermal conductivity is higher or the region is convective.

\begin{figure}[htb!]
    \centering
    \includegraphics[width=0.37\linewidth]{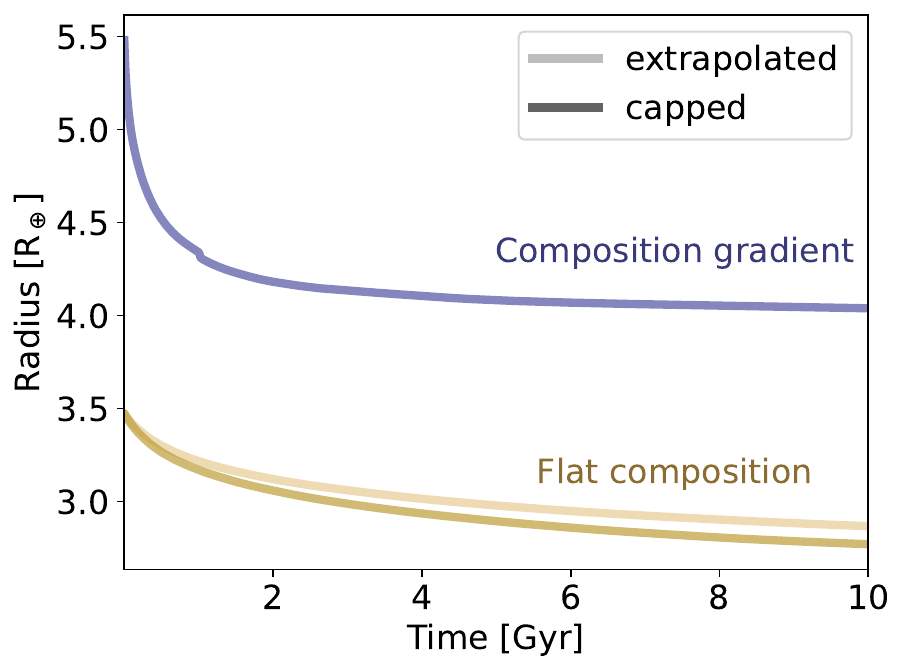}
    \caption{Radius evolution as a function of time for a $M\low{p}=10\,\ME$ planet, comparing capped and extrapolated radiative opacities. The blue lines represent a model with a composition gradient, while the orange lines show a model with flat composition profile. Both models assume a bulk heavy-element mass fraction of $Z=95$. Lighter shades indicate the evolution using the extrapolated opacity method and darker shades indicate the evolution using the capped opacity method for out of table values.}
    \label{fig:RadiusEvolutionOpacityExtrapolation}
\end{figure}
In Figure \ref{fig:RadiusEvolutionOpacityExtrapolation} we show the radius evolution for two different composition profiles. One model uses a flat composition profile, while the other uses the wide composition gradient. For both models we assume a bulk heavy-element mass fraction of $Z=0.95$ and mass of $M_\text{p}=10\,\ME$. Note, the initial central entropy is different for the two composition profiles. After 10 billion years the difference in radius is a few percent because of the different extension method of the radiative opacity tables.

\FloatBarrier
\section{Comparison of different composition gradient widths}\label{app:GradientWidth}
\begin{figure}[hbt]
    \centering
    \includegraphics[width=0.8\linewidth]{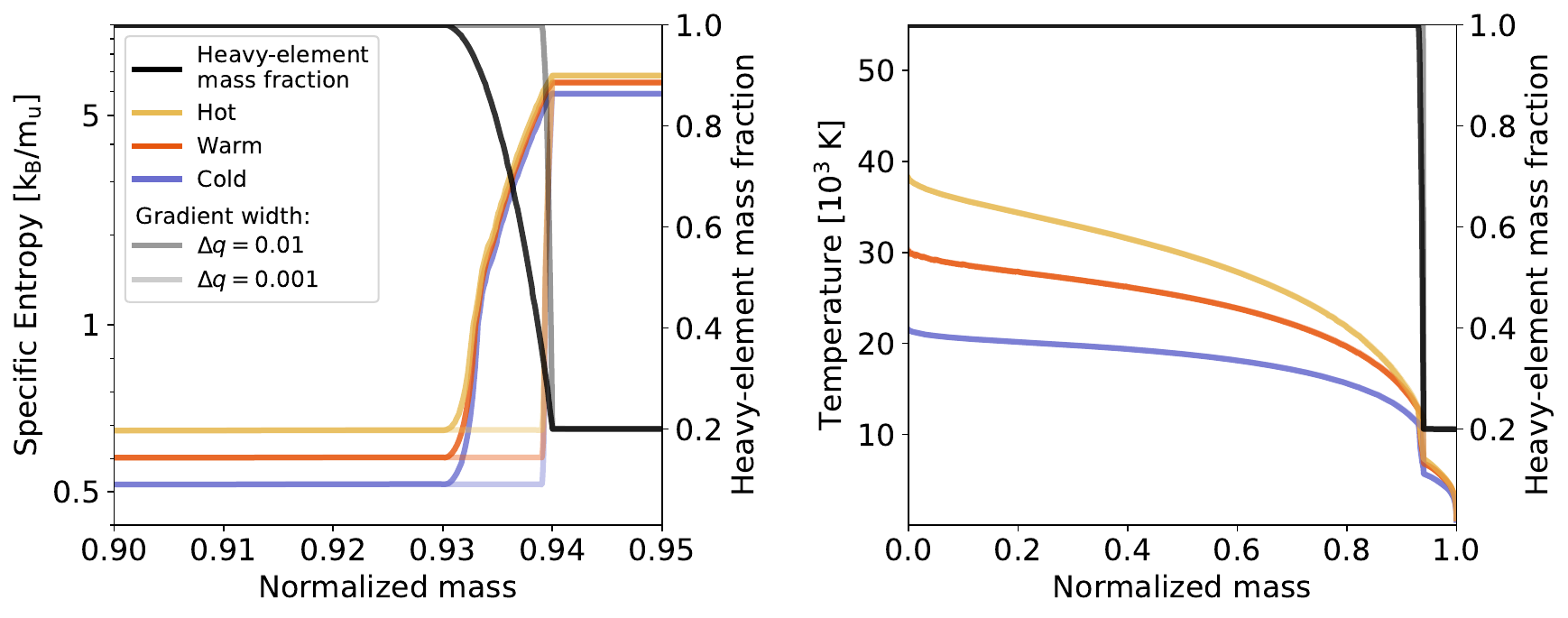}
    \caption{Initial profiles for the medium and narrow composition gradients, showing specific entropy (left) and temperature (right) as a function of normalized mass. The heavy-element mass fraction is overlaid in black in both panels, with its axis on the right. The colors blue (cold), orange (warm), and yellow (hot) correspond to different primordial entropies. The darker ($\Delta q=0.01$) and lighter ($\Delta q=0.001$) shades indicate different transition widths.}
    \label{fig:InitialEntropyCompostionAndTemperatureProfilesNarrow}
\end{figure}
\begin{figure}[hbt]
    \hspace{1.6cm}
    \includegraphics[width=0.772\linewidth]{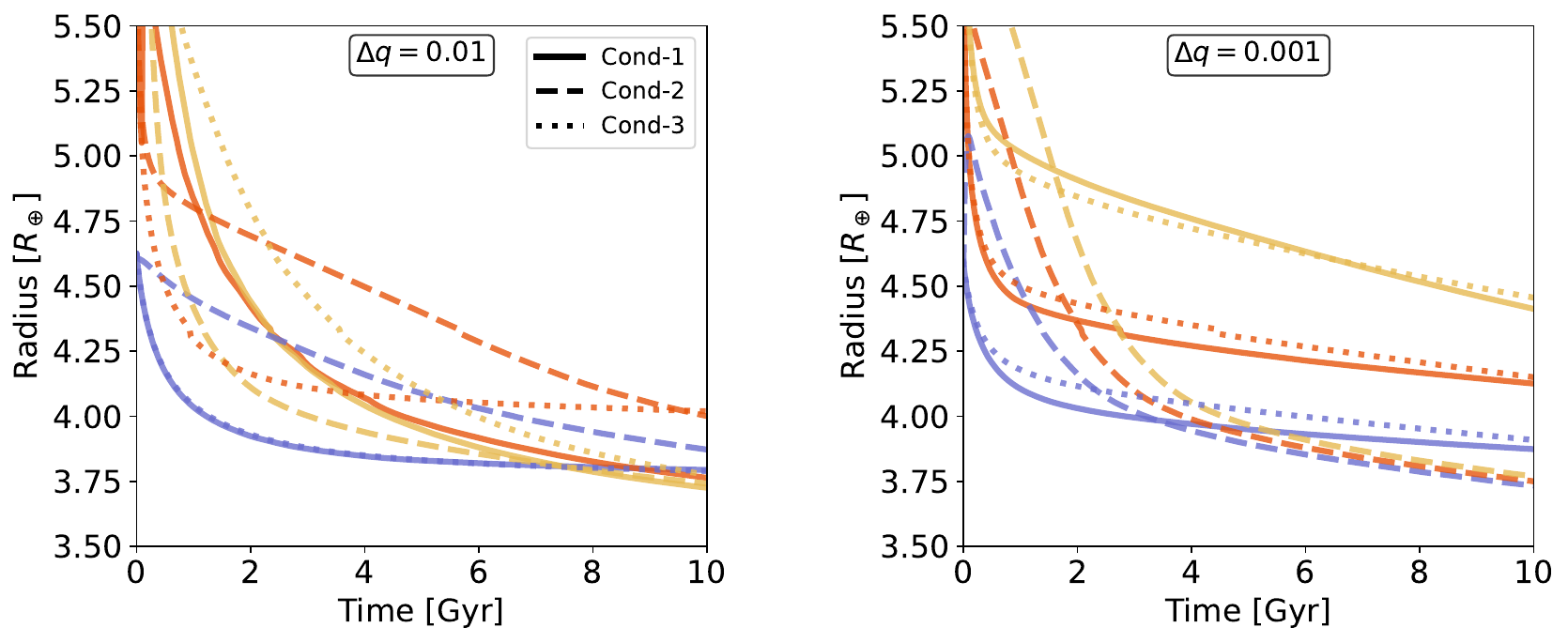}
    \caption{Radius evolution for a planet with $M\low{p}=10\,\ME$ with the medium (left) and narrow (right) composition gradient. The blue, orange, and yellow lines represent the cold, warm, and hot initial entropies, respectively. Solid lines correspond to the Cond-1 model, dashed lines to the Cond-2 model, and the dotted lines to the Cond-3 model.}
    \label{fig:RadiusEvolutionMidAndNarrow}
\end{figure}
Figure \ref{fig:InitialEntropyCompostionAndTemperatureProfilesNarrow} shows the initial entropy, heavy-element mass fraction, and temperature profiles for planets with a medium ($\Delta q = 0.01$) and narrow ($\Delta q = 0.001$) composition gradients. The narrow composition gradient is similar to the narrow composition profile presented in Paper I. Figure \ref{fig:RadiusEvolutionMidAndNarrow} shows the corresponding radius evolution. We find that a thinner transition width creates a steeper composition gradient, which increases stability against convection (see eq. \ref{eq:Ledoux}).
For the medium composition gradient ($\Delta q = 0.01$), we find that no staircase appears for a cold start. For the hot start, all the assumed conductivities lead to the formation staircases and eventually, the planets cool down to have similar radii.
For the narrow composition gradient ($\Delta q = 0.001$), the composition gradient is too steep to be eroded by convective mixing regardless of the initial entropy or thermal conductivity model. Therefore, the thermal evolution is the same as in Paper I. For Cond-2 the conductivity is sufficiently high to effectively transport the thermal energy from the deep interior to the outer envelop.  In the cases of Cond-1 and Cond-3, the thermal energy is released over a longer period of time. 

\FloatBarrier
\section{Radius evolution for different masses}\label{app:RadiusEvolutionM05M15}
\begin{figure}[hbt]
    \hspace{1.6cm}
    \includegraphics[width=0.772\linewidth]{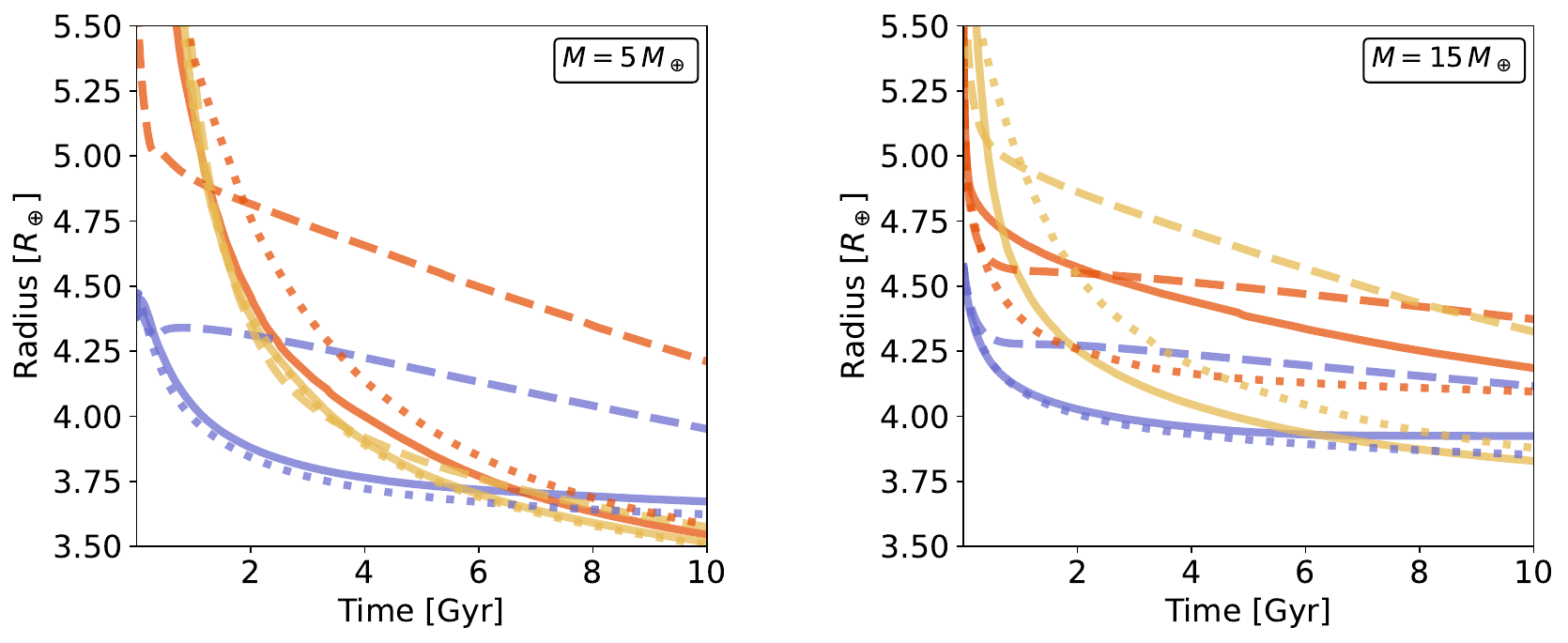}
    \caption{Radius evolution for planets with $M\low{p}=5\,\ME$ (left) and $M\low{p}=15\,\ME$ (right). The blue, orange, and yellow lines represent the cold, warm, and hot initial entropies, respectively. Solid lines correspond to the Cond-1 model, dashed lines to the Cond-2 model, and the dotted lines to the Cond-3 model.}
    \label{fig:RadiusEvolutionM05M15}
\end{figure}

Figure \ref{fig:RadiusEvolutionM05M15} shows the radius evolution for a planet with $M\low{p}=5\,\ME$ and $M\low{p}=15\,\ME$.

\FloatBarrier
\section{Final Composition Profiles}\label{app:FinalCompositionProfiles}
\begin{figure}[hbt]
    \centering
    \includegraphics[width=0.94\linewidth]{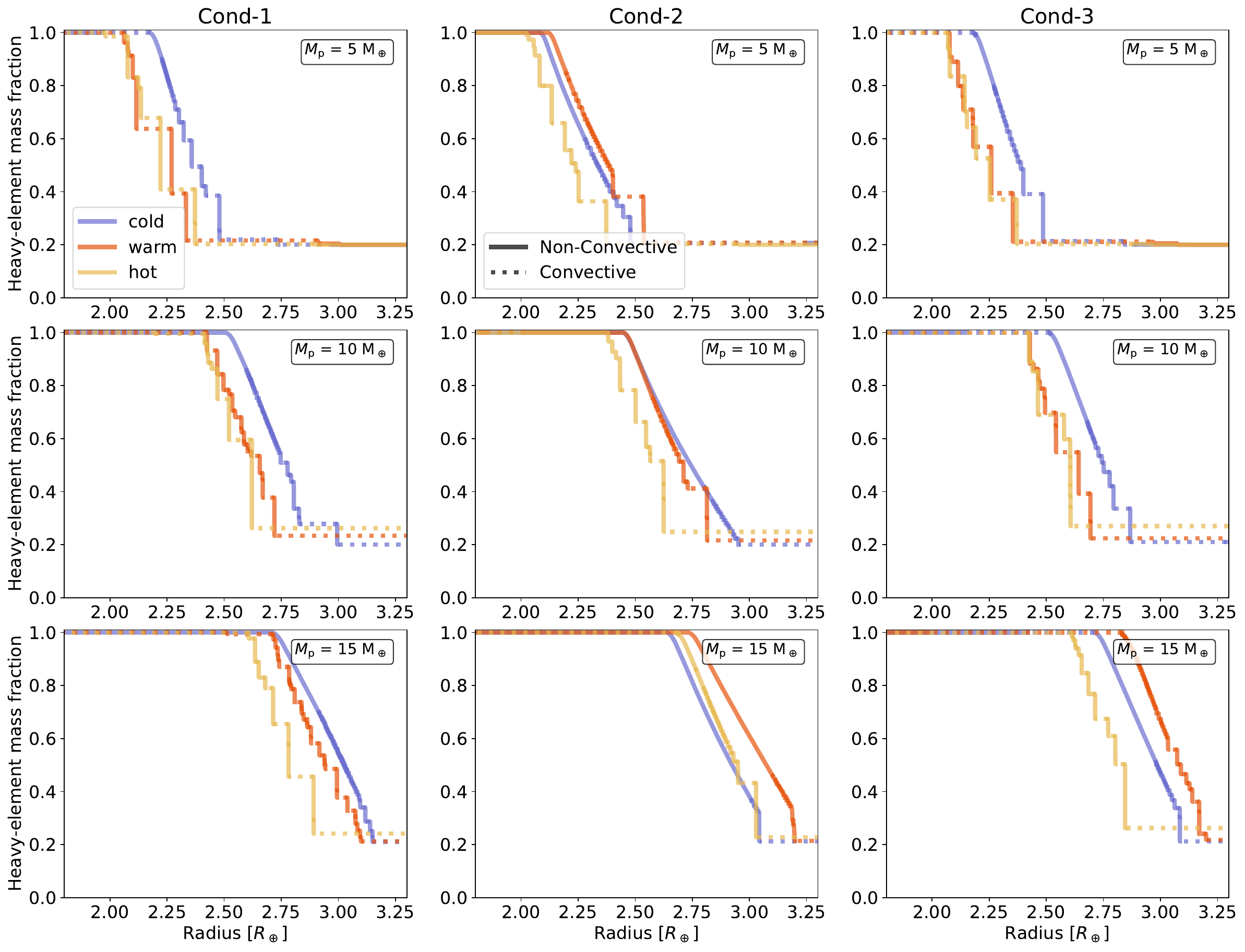}
    \caption{Heavy-element mass fraction as a function of radius at $t=10\,$Gyr for the planet with $M\low{p}=5\,\ME$ (first row), $M\low{p}=10\,\ME$ (second row), and $M\low{p}=15\,\ME$ (third row). The three columns show the three conductivity models Cond-1 (left), Cond-2 (middle), and Cond-3 (right). Different colors indicate the different initial entropies cold (blue), warm (orange), and hot (yellow). Solid lines represent non-convective regions and dotted lines represent convective regions.}
    \label{fig:FinalCompositionProfiles}
\end{figure}
Figure \ref{fig:FinalCompositionProfiles} shows the final composition profile for the wide composition gradient.

\FloatBarrier
\section{Double diffusive regions}\label{app:SemiConvZonesExtended}
\begin{figure}[hbt]
    \centering
    \includegraphics[width=0.8\linewidth]{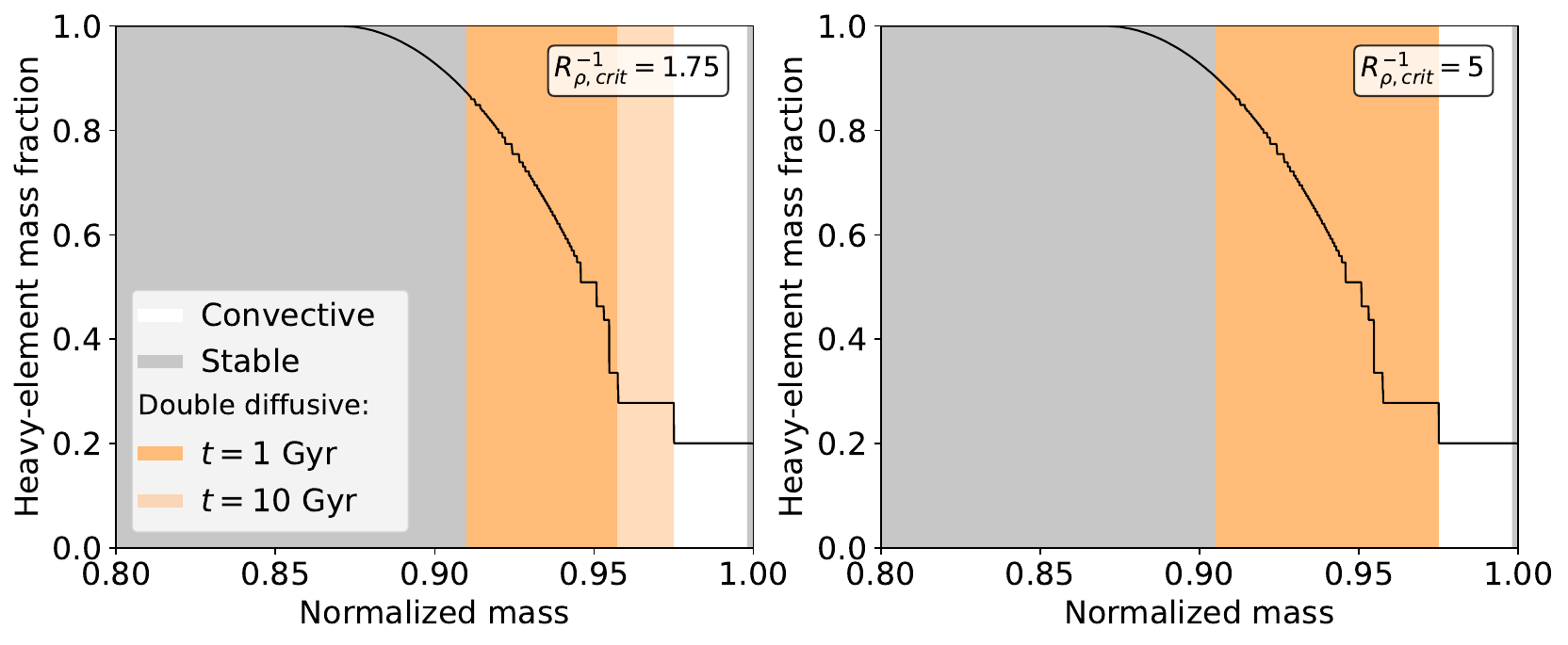}
    \caption{Same as Figure \ref{fig:Semiconvection} but with $R_{\rho,\text{crit}}^{-1}=1.25$ (left), and 5 (right) for $t=1$ Gyr (dark orange) and $t=10$ Gyr (light orange). Note that our models do not consider the onset of DDC. }
    \label{fig:SemiConvectionExtended}
\end{figure}
Figure \ref{fig:SemiConvectionExtended}: same as Figure \ref{fig:Semiconvection} but assuming  different values of $R_{\rho,\text{crit}}^{-1}$.

\end{appendix}
\end{document}